\documentclass[journal=jacsat,manuscript=article]{achemso}

\usepackage{amsmath,tikz}
\usepackage{atbegshi} 
\setkeys{acs}{articletitle = true}
\usepackage{chemformula}
\usepackage{algorithm}
\usepackage{algpseudocode}
\usepackage{braket}
\usepackage{calrsfs}
\usepackage{array}

\def\DD{\mathcal{D}}
\def\t{{\text{target}}}
\usepackage{amssymb}
\usepackage{float}
\usepackage{multirow}
\usepackage{pifont}
\usepackage{pifont}% http://ctan.org/pkg/pifont
\usepackage[x11names,table]{xcolor}

\usepackage{bm}
\addtocounter{page}{0}
\usepackage{lineno}

\begin{tocentry}
	\includegraphics[width=8cm]{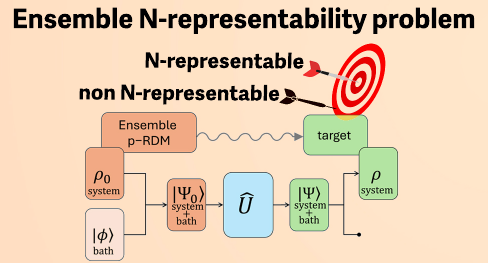}
\end{tocentry}

\title{Determining the ensemble $N$-representability of Reduced Density Matrices}

\author{Ofelia B. O\~{n}a}
\affiliation{Instituto de Investigaciones Fisicoqu\'imicas Te\'oricas y Aplicadas, Universidad Nacional de La Plata, Consejo Nacional de Investigaciones Cient\'ificas y T\'ecnicas. Diag. 113 y 64 (S/N), Sucursal 4, CC 16, 1900 La Plata, Argentina}

\author{Gustavo E. Massaccesi}
\affiliation{Departamento de Ciencias Exactas, Ciclo B\'asico Com\'un, Universidad de Buenos Aires, Ciudad Universitaria, 1428 Buenos Aires, Argentina}
\alsoaffiliation{Instituto de Investigaciones Matem\'aticas ``Luis A. Santal\'o'' (IMAS), Consejo Nacional de Investigaciones Cient\'ificas y T\'ecnicas, Universidad de Buenos Aires. Ciudad Universitaria, 1428 Buenos Aires, Argentina}

\author{Pablo Capuzzi}
\affiliation{Universidad de Buenos Aires, Facultad de Ciencias Exactas y Naturales, Departamento de F\'isica. Ciudad Universitaria, 1428 Buenos Aires, Argentina}
\alsoaffiliation{CONICET - Universidad de Buenos Aires, Instituto de F\'isica de Buenos Aires (IFIBA). Ciudad Universitaria, 1428 Buenos Aires, Argentina}

\author{Luis Lain}
\affiliation{Department of Physical Chemistry, Faculty of Science and Technology, University of the Basque Country. PO Box 644, E-48080 Bilbao, Spain}

\author{Alicia Torre}
\affiliation{Department of Physical Chemistry, Faculty of Science and Technology, University of the Basque Country. PO Box 644, E-48080 Bilbao, Spain}

\author{Juan E. Peralta}
\email{juan.peralta@cmich.edu}
\affiliation{Department of Physics, Central Michigan University, Mount Pleasant, MI, 48859, USA}

\author{Diego R. Alcoba}
\email{dalcoba@df.uba.ar}
\affiliation{Universidad de Buenos Aires, Facultad de Ciencias Exactas y Naturales, Departamento de F\'isica. Ciudad Universitaria, 1428 Buenos Aires, Argentina}
\alsoaffiliation{CONICET - Universidad de Buenos Aires, Instituto de F\'isica de Buenos Aires (IFIBA). Ciudad Universitaria, 1428 Buenos Aires, Argentina}

\author{Gustavo E. Scuseria}
\email{guscus@rice.edu}
\affiliation{Department of Chemistry, Rice University, Houston, TX 77005-1892}
\alsoaffiliation{Department of Physics and Astronomy, Rice University, Houston, TX 77005-1892}

\keywords{$N$-representability, ensemble state,  reduced density matrix, variational approach, quantum algorithm}

\begin{document}

%\linenumbers % remove for unmarked version !!!

\begin{abstract}

The $N$-representability problem for reduced density matrices remains a fundamental challenge in electronic structure theory. 
Following our previous work that employs a  unitary-evolution algorithm based on an adaptive derivative-assembled pseudo‑Trotter variational quantum algorithm to probe {\em pure-state} $N$-representability of reduced density matrices [J. Chem. Theory Comput. 2024, 20, 9968], in this work 
we propose a practical framework for  determining the {\em ensemble} $N$-representability of a $p$-body matrix. 
This is  accomplished using  a purification strategy consisting of embedding an ensemble state into a pure state defined on an extended Hilbert space, such that the reduced density matrices of the purified state reproduce those of the original ensemble. 
By iteratively applying variational unitaries to an initial purified state, the proposed algorithm minimizes the Hilbert-Schmidt distance between its $p$-body reduced density matrix and a specified target $p$-body matrix, which serves as a measure of the $N$-representability of the target.
This methodology facilitates both error correction of defective ensemble reduced density matrices, and  quantum-state reconstruction on a quantum computer, 
offering a route for density-matrix refinement. 
We validate the algorithm with numerical simulations on systems of two, three, and four electrons in both, simple models as well as  molecular systems at finite temperature, demonstrating its robustness.

\end{abstract}

\section{Introduction}

The significance of the $N$-representability problem in electronic structure theory is both foundational and far-reaching.\cite{Stillinger.book} 
At the heart of this challenge lies the quest to determine the necessary and sufficient conditions that a $p$-body reduced density matrix ($p$-RDM) must satisfy to be derivable from a $N$-electron wave function or mixed (ensemble) state.\cite{Mazziotti.book.2007,Coleman.book.2000,Coleman.RevModPhys.1963,Garrod.JMP.1963,Kummer.JMP.1967,Erdahl.IJQC.1978,Erdahl.book.1987,Mazziotti.PRL.2012}
The special case of the 2-RDM is essential for the development of quantum chemistry computational methods.\cite{Coleman.book.2000,Mazziotti.book.2007,DePrince.2024} 
Equally important, though often less emphasized, is the role of the 1-RDM in practical applications, such as  
 in Hartree-Fock  and density (matrix) functional theory, where the 1-RDM serves as a cornerstone for calculating observable properties.\cite{Kryachko1992,Parr1994,Piris2018}
Despite its theoretical interest, the $N$-representability problem presents formidable computational challenges. It belongs to the quantum Merlin-Arthur-complete class of problems: A quantum analog of the well-known nondeterministic polynomial-time complete class in classical complexity theory.\cite{Liu.PRL.2007} This classification implies that, in general, verifying whether a given $p$-RDM corresponds to a physical $N$-electron state is computationally intractable, especially as system size increases.
Recent developments have led to sets of approximate $N$-representability conditions that, while not exhaustive, are sufficiently robust to enable meaningful calculations for strongly correlated electron systems, particularly in small to medium-sized molecules.\cite{Poelmans2015,Alcoba2018b,Alcoba2018,Alcoba2020,Mazzioti2020,Mazziotti.2023,DePrince.2024} 
Moreover, Coleman\cite{Coleman.RevModPhys.1963}  and  Klyachko\cite{Klyachko,Altunbulak.CMP.2008}  derived the  necessary and sufficient $N$-representability conditions  that must be fulfilled by the 1-RDM  to be derivable from an ensemble or pure $N$-fermion quantum state, respectively. 
These conditions, which constrain the eigenvalues of the 1-RDM,  offer a   mathematically rigorous  framework for identifying physically valid fermionic occupation numbers.\cite{Altunbulak.CMP.2008,Mazziotti.book.2007,Coleman.IJQC.1978,Schilling.2013,Chakraborty.2014,Theophilou.2015,Schilling.2018,Smart.2019,Boyn.2019,Avdic.2023,PhysRevA.88.022508}

In recent work, we introduced an algorithm\cite{Massaccesi__jctc_2024} designed to numerically determine whether a given $p$-body reduced density matrix ($p$-RDM) is pure $N$-representable, or in other words, whether it can be obtained by tracing out $(N-p)$ degrees of freedom from a pure $N$-body quantum state (i.e., a single wave function or pure-state density matrix). 
The proposed algorithm operates by guiding the unitary evolution of 
an ansatz wave function through its corresponding $p$-RDM  towards a target matrix (alleged $p$-RDM), with the goal of minimizing their Hilbert-Schmidt distance (a measure of how close two matrices are in Hilbert space). 
The evolution (driven by stochastic sampling in that work) is inspired by the adaptive derivative-assembled pseudo-Trotter (ADAPT) framework, originally developed for the variational quantum eigensolver (VQE) problem,\cite{Grimsley.NatComm.2019} 
and other approaches related to it, including the contracted quantum eigensolver.\cite{Smart.2021,Warren.2024,Yordanov.2021,Tang.2021,Liu.2022,Stein.2022,Ziems.2024,Vaquero.2024}
As such, the algorithm is suited for implementation on  quantum devices.\cite{Grimsley.NatComm.2019,Fedorov.2022,Bharti.2022} 
We refer to this approach as pure ADAPT-VQA (Variational Quantum Algorithm), and its utility extends beyond mere classification. Not only can it determine whether a given matrix is a valid pure $p$-RDM, but it can also correct matrices that fail to meet the $N$-representability criteria. 
This dual capability makes the pure ADAPT-VQA a powerful tool for quantum simulations, offering a pathway to enforce physical constraints in quantum algorithms and potentially improve the accuracy of quantum chemical calculations.

In this work, we address a critical extension of the $N$-representability problem, namely, its formulation for ensemble $p$-RDMs.\cite{Coleman.book.2000,Mazziotti.book.2007} While part of the  existing literature focuses on pure-state representability,\cite{Altunbulak.CMP.2008,Mazziotti.PRA.2016,Mazziotti.book.2007,Coleman.IJQC.1978,Schilling.2013,Chakraborty.2014,Theophilou.2015,Schilling.2018,Smart.2019,Boyn.2019,Avdic.2023} 
ensemble states are equally important, especially in contexts involving thermal mixtures, open quantum systems, and statistical ensembles.\cite{Coleman.book.2000,Mazziotti.book.2007}
To address this challenge, we employ a purification-based method in which the ensemble state is represented as a pure state within an expanded Hilbert space,\cite{nielsen2000schmidt,Hughston1993} which has been  recently applied to finite temperature electronic systems through the  thermofield theory.\cite{harsha2019,harsha2020,harsha2023}
This method ensures that the reduced density matrices of both the ensemble and its purified counterpart are identical, allowing us to leverage techniques developed for pure states. Specifically, we apply the pure ADAPT-VQA algorithm introduced in our previous work,\cite{Massaccesi__jctc_2024} which enables the unitary evolution of the purified state to minimize the Hilbert-Schmidt distance between its $p$-RDM and a given target matrix.
Building on this foundation, we introduce a new algorithm, the ensemble ADAPT-VQA, which extends the capabilities of its pure-state predecessor. This hybrid framework allows us not only to determine whether a $p$-body matrix is $N$-representable, but also to discern the nature of its origin: whether it arises from a pure quantum state or from an ensemble mixture of pure states. Such a distinction is crucial for interpreting quantum simulations and for enforcing physical constraints in variational algorithms.
To validate the performance and versatility of ensemble ADAPT-VQA, we conduct numerical experiments on both model and molecular systems. Specifically, we test the algorithm on the 1-RDM and 2-RDM of four-electron systems at zero temperature, as well as on the \ch{H2} and \ch{H3} molecules at finite temperature. These examples demonstrate the  ability of the algorithm to handle pure and thermal-state scenarios, highlighting its potential for broader applications in quantum chemistry and quantum information.

\section{Theoretical Considerations}
\label{theory}

Let us consider  a  quantum mixed (ensemble) state,  $\rho$, defined  on a finite-dimensional  Hilbert space \(\mathcal{H}_s\), which can be expressed  as
\begin{equation} 
\rho = \sum_i p_i \ket{\phi_i}\bra{\phi_i}\,,
\end{equation}
where  \(\ket{\phi_i} \in \mathcal{H}_s\) are pure-state wave functions, and $p_i$
probabilities satisfying \(p_i \ge 0\) with \(\sum_i p_i = 1\).
According to the  Schr\"odinger-Hughston-Jozsa-Wootters theorem,\cite{Kirkpatrick2006} 
any such mixed state \(\rho\) admits a purification: There exists an auxiliary finite-dimensional Hilbert space \(\mathcal{H}_b\) and a pure-state wave function $\ket{\Psi_{sb}} \in \mathcal{H}_{sb}$ (with  $\mathcal{H}_{sb} = \mathcal{H}_s \otimes \mathcal{H}_b$ an extended Hilbert space), referred to as a purification of \(\rho\), such that\cite{nielsen2000schmidt}
\begin{equation}
\label{eq:2}
\rho = \mathrm{Tr}_b \bigl( \ket{\Psi_{sb}}\bra{\Psi_{sb}} \bigr) \,,
\end{equation}
\noindent
where $\mathrm{Tr}_b $ implies the complete trace over the degrees of freedom on the auxiliary Hilbert space $\mathcal{H}_b$.
Moreover, every purification of \(\rho\) can be written in the form
\begin{equation}
\label{eq:3}
\ket{\Psi_{sb}} = \sum_i \sqrt{p_i}\,\ket{\phi_i} \otimes \ket{b_i}\,,
\end{equation}
where \(\{\ket{b_i}\}_i\) is an orthonormal basis of \(\mathcal{H}_b\).  
Thus, one can use this strategy to embed an ensemble state into a pure state  defined in an extended Hilbert space,  such that the $p$-RDMs of the purified  ensemble are reproduced by those of the pure state.

Eq.~\ref{eq:3} provides a practical approach to numerically
determine the ensemble $N$-representability of a $p$-RDM by unitarily evolving a pure state. This can be accomplished  using an  algorithm that has been recently introduced by us to deal with the pure $N$-representability problem.\cite{Massaccesi__jctc_2024}
Here we summarize  the newly proposed algorithm employed for the ensemble case.  
The underlying
idea is to find a sequence of unitary transformations that evolve an
initial $2N$-particle wave function $\ket{\Psi_0}$ built according
to Eq.~\ref{eq:3}, e.g., 
\begin{equation}
\label{eq2b}
\ket{\Psi_0}=  \ket{\phi_0} \otimes \ket{\phi_0}  \,,
\end{equation}
\noindent
with $\ket{\phi_0}$ holding $N$ particles. 
The evolution is such that 
 the Hilbert-Schmidt distance $\DD$ between the $p$-RDM corresponding to the  evolved state $\ket{\Psi_k}$, 
 \begin{equation}
  ^p \rho_k =   \,p!\binom{N}{p}    \mathrm{Tr}_{N-p} \bigl( \mathrm{Tr}_b \bigl( \ket{\Psi_{k}}\bra{\Psi_{k}} \bigr) \bigr)\,,
 \end{equation}
  and a given (possibly) ensemble $N$-representable target
$p$-RDM, $^p\rho_\t$, is minimized. 
The (squared) distance is calculated according to the Hilbert-Schmidt norm, 
\begin{equation}
    \label{eq:Distance}   
\DD_k  =  \|^p\rho_k-\,^p\rho_{\t}\|_{2}^2  \,.
\end{equation}
At convergence, the ADAPT-VQA yields a minimum feasible distance for the problem at hand, $\DD_{\min}$, and the state  $\ket{\Psi_{\min}}$.
Thus, 
if  the algorithm evolves the distance $\DD_k$ to numerical  zero, then 
$^p\rho_{\t}$ is ensemble $N$-representable. 
In general, $\DD_{\min}$ can be considered as a measure of the degree of ensemble $N$-representability of $^p\rho_\t$, and the evolved $^p\rho$ is an ensemble  $N$-representable (physical) reduced state that is closest to $^p\rho_\t$. 
Hence, the algorithm allows to correct the possible error associated with the lack of ensemble $N$-representability of $^p\rho_\t$.

The minimization algorithm employed in this work  is schematized in Algorithm~\ref{Scheme}. The procedure follows the schemes  previously introduced by us to determine the $N$-representability of
pure RDMs and transition RDMs.\cite{Massaccesi__jctc_2024,10.1063/5.0275683} In the present work, we adopt the same  gradient-based strategy  utilized for  transition RDMs, in combination   with an  antihermitian excitation-deexcitation  operator pool $\{\hat{O}\}$ suitable for fermions,\cite{Mazziotti.pra.2007,OMalley.PRX.2016,Barkoutsos.PRA.2018,Evangelista.JCP.2019}
\begin{align}
  \label{eq:fermionic}
  \hat{O}_{i k} &= \hat{a}_{i}^{\dagger}\hat{a}_{k}-\hat{a}_{k}^{\dagger} \hat{a}_{i}, \; \; 
\end{align}
and
\begin{align}
  \label{eq:fermionic2}
  \hat{O}_{i j k l} &= \hat{a}_{i}^{\dagger} \hat{a}_{j}^{\dagger} \hat{a}_{k} \hat{a}_{l}
                -\hat{a}_{k}^{\dagger} \hat{a}_{l}^{\dagger} \hat{a}_{i} \hat{a}_{j}\,,
\end{align}
\noindent for one- and two-particle excitations, respectively. In Eqs.~\ref{eq:fermionic} and \ref{eq:fermionic2},  $\hat{a}^{\dagger}$ and $\hat{a}$ are the usual fermionic creation and annihilation operators acting on an
orthonormal finite single-particle basis associated to the extended Hilbert space $\mathcal{H}_{sb}$. Furthermore, since we opt to work on a canonical ensemble framework,   the operator pool is restricted in such a way that no electronic transitions changing the number of particles in  the system of interest nor in the auxiliary system are allowed. 
In practice, the operators in Eqs.~\ref{eq:fermionic} and \ref{eq:fermionic2} are represented as a combination of    Pauli operators  via the  Jordan-Wigner transformation,\cite{Jordan1993} 
ensuring that the proposed algorithm 
is  suitable for simulations on  quantum computers.\cite{Tilly.2022}

\begin{figure}
  \begin{algorithm}[H]
  \caption{Distance ($\DD$) minimization algorithm}
    \label{Scheme}
    \begin{algorithmic}
    \Require Target $^p\rho_\t$ and initial state $\ket{\Psi_0}$ defined over an extended Hilbert space  $\mathcal{H}_{sb}$.
    \Ensure $\DD_\text{min}$, $\ket{\Psi_\text{min}}$.
   \State Prepare operator pool $\{\hat{O}\}$ corresponding to  $\mathcal{H}_{sb}$.
   \Procedure{}{}
   \While{$k<k_\text{max}$ and $|\DD_k-\DD_{k-1}|>\delta$}
        \For{$\hat{O} \in \{\hat{O} \}$}
            \State Minimize  distance to the target $^p\rho_\t$ for $\ket{\Psi}= e^{\theta\,\hat{O}} \ket{\Psi_{k-1}}$ with respect to $\theta$.
        \EndFor
    \State Keep $\hat{O}_{\text{min}}$ and the corresponding $\theta_{\text{min}}$ that give the minimum distance $\DD$. 
    \State The new state is $\ket{\Psi_k} =  e^{\,\theta_\text{min}\,\hat{O}_\text{min}}  \ket{\Psi_{k-1}}$.
    \State $\DD_k \gets \DD_\text{min}$ , $\ket{\Psi_k} \gets  \ket{\Psi_\text{min}}$.
    \State $k \gets k+1$.
    \EndWhile
    \EndProcedure
    \end{algorithmic}
\end{algorithm}
\end{figure}

\newpage

\section{Computational Details}

The implementation of the algorithm described in the previous Section utilizes an in-house code written in {\sc Python} that uses  {\sc OpenFermion}\cite{OpenFermion2020} for the  Jordan–Wigner mapping and the {\sc OpenFermion} module of {\sc PySCF}\cite{pyscf_wires,pyscf_jcp} for integral computation and manipulation. 
Computations reported in this work are carried out on a simulated noiseless quantum device. 
 The original ensemble $N$-electron problem is embedded into a extended  pure $2N$-electron problem (Eq.~\ref{eq:3}). 
 The former uses $K$ spatial orbitals as single-particle basis, while the latter uses $2K$ orbitals. 
For the 4-electron model systems at zero temperature, we use $K=3$ and $K=4$ single-particle spatial orbitals, 
referred to as $(4e,3o)$ and  $(4e,4o)$ hereafter,
while for the 
 \ch{H2} and \ch{H3} molecular systems at finite temperature we use  STO-3G basis set ($K=2$ and $K=3$, respectively). This choice allows us to deal with  1- and 2-RDMs,  which are commonly the most relevant for electronic structure problems, while  keeping the computations tractable.
 This leads to a pool of 378 and 1196  operators for the $(4e,3o)$, $(4e,4o)$, respectively, while for the  \ch{H2} and \ch{H3} molecular cases the number of operators in the pool is 72, and 378, respectively.
The convergence threshold, $\delta$, in the iterative process (Algorithm~\ref{Scheme}) is varied within the range $[5\times10^{-9}, 3\times10^{-5}]$. A smaller $\delta$ is used for stricter convergence requirements, such as for $N$-representable one-body targets, while $\delta$ is increased for two-body targets and in cases of a stronger $N$-representability violation. 
Details of the number of iterations to reach convergence are available in the supporting information (SI).

\section{Results and Discussion}

\subsection{Model systems at zero temperature}

Given a $p$-RDM, one can determine if it is pure $N$-representable utilizing the previously proposed pure ADAPT-VQA. If the $p$-RDM results non pure $N$-representable, one can utilize the ensemble ADAPT-VQA proposed in this work to determine  its ensemble $N$-representability.  
Thus, the two algorithms serve to classify a $p$-RDM as  pure or ensemble $N$-representable.
We first test this idea to numerically determine if the  $1$- and  $2$-RDMs  of  two  analytical model systems are pure or  ensemble $N$-representable. 
To assess the validity of our numerical results, we make use of Klyachko’s $N$-representability inequalities (also known as generalized Pauli conditions), which provide a set of necessary and sufficient analytical constraints that  pure $1$-RDMs must satisfy.\cite{Klyachko,Altunbulak.CMP.2008} 
For the $(4e,3o)$ and  $(4e,4o)$ model systems, we construct target $p$-RDMs as 
\begin{equation}
^p\rho_\t\,=\,p!\binom{N}{p}\mathrm{Tr}_{N-p}(\rho)  \,,
\end{equation}
where the state $\rho$ is  a convex linear  combination of two pure states, $\rho_{1}$ and $\rho_{2}$ (shown in Table~\ref{tab:4e3o}),
\begin{equation}
\label{eq:omega}
    \rho = w\; {\rho}_{1} + (1-w) \; {\rho}_{2}  \qquad 0\leq w \leq 1  \,.
\end{equation}  
The parameter $w$ in Eq.~\ref{eq:omega} is set to a value of $0.0$ or $0.5$ for $\rho$  representing either a pure, or mixed (ensemble) state, respectively. This choice is  such that
Klyachko's conditions in the $(4e,3o)$ model system are satisfied for both $w=0.0$ and $w=0.5$, whereas  in   the $(4e,4o)$ case, these conditions are satisfied only for $w=0.0$ (see Table S1 in the SI for the complete list of inequalities). 
Table~\ref{tab:Klyachko} shows our results for the 1-RDM of the $(4e,3o)$ and $(4e,4o)$ model systems for $w=0.0$ and $w=0.5$. 
In all cases, the initial state is taken as $\rho_1$ for the pure case, or its purification  for the ensemble case. 
For the $(4e,3o)$ model system, the pure and ensemble ADAPT-VQA yield converged $\DD_{\min}$ that are numerically zero for both $w$ values, indicating that there exists pure states that are compatible with the target 1-RDMs. This is consistent with the satisfaction of all Klyachko's conditions, as shown in  Table~\ref{tab:Klyachko}. 
For the $(4e,4o)$ model system, on the other hand, both the pure and ensemble ADAPT-VQAs yield 
converged $\DD_{\min} \approx 0$ for $w=0.0$, while for $w=0.5$ 
(for which Klyachko's conditions are not fulfilled), $\DD_{\min}$ approaches  zero only for the ensemble algorithm. This highlights the ability of the algorithms to discern the pure or ensemble $N$-representability  nature   of 1-RDMs. 
Table~\ref{tab:Klyachko} also reports the number of Klyachko's inequalities that are satisfied for each case.
%Table~\ref{tab:Klyachko} also list all Klyachko's inequalities that are , highlighting those that are not satisfied for the $w=0.5$ case.
We note that all targets are identified as ensemble $N$-representable by the ensemble ADAPT-VQA, in agreement with 
the fulfillment of Coleman's ensemble $N$-representability conditions (1-RDM eigenvalues between 0 and 1, and tracing to the number of electrons).\cite{Coleman.RevModPhys.1963}

%%%%%%%%%%%%%%%%%%%%%%%%%%%%%%%%%%%%%%%%%%%%%%%%%%%%%%%%%%%%%%%%%%%%
\begin{table}
    \centering
    \caption{States used to construct the target RDMs for the $(4e,3o)$ and $(4e,4o)$ model systems in Eq.~\ref{eq:2}.}
\begin{tabular}{|c|>
{\centering\arraybackslash}p{3.4cm}|>{\centering\arraybackslash}p{3.4cm}|}
\hline
     &    \multicolumn{2}{c|}{State} \\ 
         \cline{2-3}
   Model system & $\rho_1$ & $\rho_2$ \\[3pt] 
   \hline
   ($4e,3o$)      & $|1^\alpha \; 1^\beta \; 2^\alpha \; 2^\beta \rangle$ & $|1^\alpha \; 1^\beta \; 3^\alpha \; 3^\beta \rangle$ \\ [4pt]
   ($4e,4o$)      & $|1^\alpha \; 1^\beta \; 2^\alpha \; 2^\beta \rangle$ & $|1^\alpha \; 1^\beta \; 3^\alpha \; 2^\beta \rangle$ \\ [4pt]
   \hline
    \end{tabular}
    \label{tab:4e3o}
\end{table}
%%%%%%%%%%%%%%%%%%%%%%%%%%%%%%%%%%%%%%%%%%%%%%%%%%%%%%%%%%%%%%%%%%%%%%%%%%%%%%%%%

%%%%%%%%%%%%%%%%%%%%%%%%%%%%%%%%%%%%%%%%%%%%%%%%%%%%%%%%%%%%%%%%%%%%%%%%%
\begin{table}
\centering
\caption{Converged $\DD_{\min}$ from the pure and ensemble ADAPT-VQA for 4-electron model systems for 1-RDMs, along with the  1-RDMs eigenvalues, 
 and the number of  Klyachko's pure $N$-representability
inequalities that is fulfilled in each case, $N_\text{ineq}$ (out of a total of 15 for this model).
%the complete list of  Klyachko's pure $N$-representability
%inequalities, and their fulfillment ($\checkmark$) or violation ($\bm{\times}$).
\label{tab:Klyachko}}
\begin{tabular}{|c|c|c|c|c|}
\hline
\multirow{3}{*}{1-RDM eigenvalues} & \multicolumn{4}{c|}{Model system} \\
\cline{2-5}
                                & \multicolumn{2}{c|}{$(4e,3o)$} & \multicolumn{2}{c|}{$(4e,4o)$} \\
\cline{2-5}
                             &  $w=0.0$ &  $w=0.5$  &  $w=0.0$ &  $w=0.5$     \\
\hline
$\lambda_1$&1.000&	 	 1.000&1.000&	 	 1.000 \\
$\lambda_2$&1.000&	 	 1.000&1.000&	 	 1.000 \\
$\lambda_3$&1.000&	 	 0.500&1.000&	 	 1.000 \\
$\lambda_4$&1.000&	 	 0.500&1.000&	 	 0.500 \\
$\lambda_5$&0.000&	 	 0.500&0.000&	 	 0.500 \\
$\lambda_6$&0.000&	 	 0.500&0.000&	 	 0.000 \\
$\lambda_7$&     &	          &0.000&	 	 0.000 \\
$\lambda_8$&     &            &0.000&		 0.000 \\
\hline
$N_\text{ineq}$  & 15 & 15& 15&  7 \\
\hline
\multirow{2}{*}{ADAPT-VQA} &    \multicolumn{4}{c|}{\multirow{2}{*}{$\DD_\text{min}{}$}}  \\
          &       \multicolumn{4}{c|}{}  \\
\hline
pure    & 0.0 & $4.89 \times 10^{-9}$ & 0.0 & $1.25 \times 10^{-1}$ \\
\hline
ensemble& 0.0 & $4.89 \times 10^{-9}$ & 0.0 & $2.45 \times 10^{-9}$\\
\hline
\end{tabular}
\end{table}

Next, we extend the analysis to  2-RDMs. 
While a set of necessary pure $N$-representability conditions has been reported for 2-RDMs,\cite{Mazziotti.PRA.2016}  the complete set of necessary and sufficient  pure $N$-representability conditions remains unknown. 
Our  ADAPT-VQA can be used to numerically decide the pure or ensemble $N$-representability of 2-RDMs. To highlight this,  we analyze the 
$N$-representability of 2-RDMs  
for the  $(4e,3o)$ and  $(4e,4o)$ model systems. Our results are summarized in Table~\ref{tab:3-2rdm}. 
For these model systems, the algorithm identifies all the 2-RDMs  as ensemble $N$-representable, but only the cases corresponding to $w=0.0$ are identified as pure   $N$-representable, as expected. 
This brings an interesting question:  Can the method be used to detect  if  an ensemble $N$-representable  $p$-RDM   would  contract to a pure $N$-representable $q$-RDM ($q<p$)? 
To answer this question, we build target RDMs for a $(4e,4o)$ model system from an ensemble state as linear combination of two pure states (Eq.~\ref{eq:omega} with $w=0.5$), which are in turn constructed as shown in Table~\ref{tab:excitation}. These states are chosen so that they differ in one, two, or three spin-orbitals. 
A summary of the results is shown in Table~\ref{tab:excitation2}.
For the case of triple substitution, the  1-RDM and the 2-RDM can be also  
obtained  from a wave function (a linear combination of those reported in Table~\ref{tab:excitation}), 
and thus they are both pure and ensemble $N$-representable.
This indicates that the $N$-electron state that generated these RDMs cannot be uniquely determined solely from them.
Thus, $\DD_{\min}$ is expected to approach zero as the pure and ensemble algorithms evolve. The numerical results shown in Table~\ref{tab:excitation2} confirm that the ADAPT-VQA yields the expected outcomes.
For the doubly substituted  case, only the pure algorithm yields  $\DD_{\min} \neq 0$ for the 2-RDM, showing that indeed this matrix corresponds to an ensemble state, 
%while its contraction leads to a pure $N$-representable 1-RDM.
while its contraction leads to a 1-RDM that can be identified as pure $N$-representable.
For completeness, Table~\ref{tab:excitation2} shows the results for the single substitution case, indicating  that  neither the target 1-RDM nor the target 2-RDM can be obtained from a pure state. 
It is important to emphasize that, although the 1- and 2-RDMs are constructed from an ensemble state, the pure algorithm is able to find a compatible wave function that yields the same RDMs (provided that $\DD_{\min}\rightarrow 0$), 
highlighting its potential use in quantum tomography.\cite{PhysRevLett.118.020401}

%%%%%%%%%%%%%%%%%%%%%%%%%%%%%%%%%%%%%%%%%%%%%%%%%%%%%%%%%%%%%%%%%%%%%%%%%%%%%%%%%
\begin{table}[h]
\centering
\caption{Converged $\DD_{\min}$ from the pure and ensemble ADAPT-VQA for the $(4e, 3o)$ and $(4e, 4o)$ model systems for 2-RDMs.\label{tab:3-2rdm} } 
\begin{tabular}{|c|>
{\centering\arraybackslash}p{3.4cm}|>{\centering\arraybackslash}p{3.4cm}|>{\centering\arraybackslash}p{3.4cm}|>{\centering\arraybackslash}p{3cm}|}
\hline
\multirow{3}{*}{Algorithm} & \multicolumn{4}{c|}{$\DD_\text{min}{}$} \\
\cline{2-5}
                                & \multicolumn{2}{c|}{$(4e,3o)$} & \multicolumn{2}{c|}{$(4e,4o)$} \\
\cline{2-5}
                             &  $w=0.0$ &  $w=0.5$  &  $w=0.0$ &  $w=0.5$     \\
\hline
pure    & 0.0                    & $2.00$   & 0.0                    & $4.25$ \\
\hline
ensemble& $1.07 \times 10^{-14}$ & $3.11 \times 10^{-12}$ & $1.07 \times 10^{-14}$ & $1.07 \times 10^{-14}$\\
\hline
\end{tabular}
\end{table}
%%%%%%%%%%%%%%%%%%%%%%%%%%%%%%%%%%%%%%%%%%%%%%%%%%%%%%%%%%%%%%%%%%%%%%%%%%%%%%

%%%%%%%%%%%%%%%%%%%%%%%%%%%%%%%%%%%%%%%%%%%%%%%%%%%%%%%%%%%%%%%%%%%%%%%%%%%%
\begin{table}[h]
    \centering
    \caption{Singly- (1), doubly- (2), and triply- (3) spin-orbital-substituted states   used to construct the target RDMs for the $(4e, 4o)$ model system. \label{tab:excitation}}
\begin{tabular}{|c|>
{\centering\arraybackslash}p{3.4cm}|>{\centering\arraybackslash}p{3.4cm}|}
    \hline
   Substitution & $\rho_1$ & $\rho_2$ \\[3pt] 
   \hline
   1      & $|1^\alpha \; 1^\beta \; 2^\alpha \; 2^\beta \rangle$ & $|1^\alpha \; 1^\beta \; 3^\alpha \; 2^\beta \rangle$ \\ 
   2      & $|1^\alpha \; 1^\beta \; 2^\alpha \; 2^\beta \rangle$ & $|1^\alpha \; 1^\beta \; 3^\alpha \; 3^\beta \rangle$ \\ 
   3      & $|1^\alpha \; 1^\beta \; 2^\alpha \; 2^\beta \rangle$ & $|1^\alpha \; 3^\beta \; 4^\alpha \; 4^\beta \rangle$ \\ 
   
   \hline
    \end{tabular}
    \label{tab:my_label}
\end{table}

%%%%%%%%%%%%%%%%%%%%%%%%%%%%%%%%%%%%%%%%%%%%%%%%%%%%%%%%%%%%%%%%%%%%%%%%%%%%
\begin{table}[h]
\centering
\caption{
Converged $\DD_{\min}$ from the pure and ensemble ADAPT-VQA for the $(4e, 4o)$  model system for 1- and 2-RDMs with $w=0.5$. The targets are the (ensemble)  reduced density matrices corresponding to the states constructed from multiple substitutions as shown in Table~\ref{tab:excitation}.  \label{tab:excitation2} }
\begin{tabular}{|c|>
{\centering\arraybackslash}p{2.5cm}|>{\centering\arraybackslash}p{2.5cm}|>
{\centering\arraybackslash}p{2.5cm}|>
{\centering\arraybackslash}p{2.5cm}|}
\hline
\multirow{4}{*}{Substitution} & \multicolumn{4}{c|}{$\DD_\text{min}{}$} \\
\cline{2-5}
                                  & \multicolumn{2}{c|}{1-RDM} & \multicolumn{2}{c|}{2-RDM} \\
\cline{2-5}
                                  & \multicolumn{4}{c|}{Algorithm} \\
\cline{2-5}
                                  &  pure &  ensemble &  pure &  ensemble     \\
\hline
1    & $1.25 \times 10^{-1}$ & $2.45 \times 10^{-9}$ & $4.25$                 &   $1.07 \times 10^{-14}$                    \\
2   & $4.89 \times 10^{-9}$ & $4.89 \times 10^{-9}$ & $2.00$                 &   $9.66 \times 10^{-9}$                    \\
3  & $2.45 \times 10^{-9}$ & $2.45 \times 10^{-9}$ & $1.42 \times 10^{-14}$ &   $2.77 \times 10^{-8}$                    \\
\hline
\end{tabular}
\end{table}
%%%%%%%%%%%%%%%%%%%%%%%%%%%%%%%%%%%%%%%%%%%%%%%%%%%%%%%%%%%%%%%%%%%%%%%%%%%%%%

So far we have tested our ADAPT-VQA with $N$-representable RDMs as targets. 
To test the ensemble algorithm with cases where the $N$-representability is violated, we have incorporated numerical noise to RDMs, $^p\rho$,  using
\begin{equation}
\label{eq:noise}
    ^p\rho_\t(\varepsilon) =\;^p\rho + \varepsilon R \,, 
\end{equation}
where $R$ is a matrix of random numbers taken from a uniform probability distribution in $[-1,1]$ and $\varepsilon$ is in  the interval $[0,0.1]$.
In Eq.~\ref{eq:noise}, $^p\rho$ is constructed according to Eq.~\ref{eq:omega} with $w=0.5$ and $\rho_1$ and $\rho_2$ given in Table~\ref{tab:4e3o}.
The values of $\varepsilon$ are chosen so that the resulting noise produces  
noticeably $N$-representability defects on $^p\rho_\t(\varepsilon)$. 
Thus, the algorithm should evolve  the initial RDM  towards the target up to a certain limit, depending on the strength of the noise. 
Larger noise strengths should lead to larger $\DD_{\min}$, while a noiseless target
should converge $\DD_{\min}\rightarrow 0$ as previously shown. 
Table~\ref{tab:noise1} shows the converged $\DD_{\min}$ for $^1\rho_\t(\varepsilon)$ and $^2\rho_\t(\varepsilon)$ corresponding to  the $(4e,4o)$ model system for different  noise strengths. 
The converged $\DD_{\min}$ in Table~\ref{tab:noise1} can then be considered as a numerical measure of the violation of the ensemble $N$-representability. 
These results are in line with the findings in Ref.~\citenum{Massaccesi__jctc_2024} for the pure ADAPT-VQA when considering systematic violations of $N$-representability conditions.  
This emphasizes applications for the ADAPT-VQA: It can be used not only to determine the ensemble $N$-representability of an alleged $p$-RDM, but  also to construct  an ensemble $p$-RDM (the evolved RDM) that is closest to the target $^p \rho_\t$.

%%%%%%%%%%%%%%%%%%%%%%%%%%%%%%%%%%%%%%%%%%%%%%%%%%%%%%%%%%%%%%%%%%%%%%%%%%
\begin{table}[H]
\centering
\caption{
Converged $\DD_{\min}$ from the ensemble ADAPT-VQA for the 1- and 2-RDMs in the $(4e,4o)$ model system. 
The targets are
constructed by adding random noise of strength $\varepsilon$ to the reduced density matrices (see text for details). 
\label{tab:noise1}}
\begin{tabular}{|c|>
{\centering\arraybackslash}p{3.4cm}|>{\centering\arraybackslash}p{3.4cm}|}
\hline
\multirow{2}{*}{$\varepsilon$} & \multicolumn{2}{c|}{$\DD_\text{min}{}$} \\
\cline{2-3}
                                & \multicolumn{1}{c|}{1-RDM} & \multicolumn{1}{c|}{2-RDM} \\
\hline
$0.0$	    &	 $2.45 \times 10^{-9}$        &  $1.07 \times 10^{-14}$ \\
\hline
$10^{-2}$	&	 $2.02 \times 10^{-3}$        &  $1.38 \times 10^{-1}$	\\
\hline
$10^{-1}$	&	 $2.19 \times 10^{-1}$	 	  & $1.33 \times 10^{1}$	\\
\hline
\end{tabular}
\end{table}

%%%%%%%%%%%%%%%%%%%%%%%%%%%%%%%%%%%%%%%%%%%%%%%%%%%%%%%%%%%%%%%%%%%%%%%%%%%%%%%%%%

%%%%%%%%%%%%%%%%%%%%%%%%%%%%%%%%%%%%%%%%%%%%%%%%%%%%%%%%%%%%%%%%%%%%%%%%%%%%%%%%
\subsection{Molecular Systems at Finite Temperature}

We next assess the usability of the ensemble ADAPT-VQA for  \ch{H2} and linear  \ch{H3} molecular systems at a finite temperature.  
In this case, 
the $p$-RDMs  are constructed from their canonical ensemble thermal states at temperature $T$,
\begin{equation}
\label{eq:Z}
    \rho = Z^{-1} \sum_i e^{-E_i/k_BT} \rho_i \,,
\end{equation}
\noindent
where $k_B$ is the Boltzmann constant,  and the canonical  partition function $Z$ is given by  
\begin{equation}
Z = \sum_i e^{-E_i/k_BT}\,.
\end{equation}
\noindent
In Eq.~\ref{eq:Z},  $E_i$ and $\rho_i$ are the eigenenergies and eigenstates of the electronic Hamiltonian $H$, which can be written in the second quantization formalism as\cite{Surjan.book.1989}
\begin{equation}
{\hat H} = \sum_{ij}\;\langle i | h |j\rangle\;{\hat a}^{\dag}_{i}{\hat a}_{j} +
\frac{1}{4}\;\sum_{ijkl}\; \langle ij| v |kl\rangle\;{\hat a}^{\dag}_{i}{\hat a}^{\dag}_{j}{\hat a}_{l}{\hat a}_{k}\,,
\label{eq:Hgen}
\end{equation}
where $\langle i| h |j\rangle$ and $\langle ij| v |kl\rangle $ are the standard
one- and two-electron antisymmetrized integrals, respectively.

We  consider both molecular systems at two nuclear configurations:  close to equilibrium  ($\mathrm{R_{HH}}$=0.75~{\AA}) and  stretched ($\mathrm{R_{HH}}$=1.5~{\AA}). 
In all cases $k_BT$ is the energy gap between the ground and first excited states. With this choice, the stretched configuration presents a thermal distribution with larger weights for the lower energy states than that of the equilibrium configuration.
Using the ensemble RDMs corresponding to these 
thermally weighted states, we  construct target matrices by adding random noise of strength $\varepsilon$ as described above (see Eq.~\ref{eq:noise}). 
We observe that for \ch{H2}, the 2-RDM is not a reduced state of the system.
In all cases, the initial state has been constructed considering Eq.~\ref{eq2b}, with $\ket{\phi_0}$
the Hartree-Fock ground state.
In Table~\ref{tab:H2H3} we show the converged $\DD_\text{min}$ for three values of $\varepsilon$. 
The results should serve as an indication of what to  expect for the ensemble ADAPT-VQA for  physical systems when the target matrices deviate from ideal ensemble $N$-representability. 
For example, when this deviation increases, as measured by the parameter $\varepsilon$,   $\DD_\text{min}$ also increases to values of $\sim10^{-2}$ and $\sim10^0$ for the 1- and 2-RDM for   $\varepsilon=0.1$, respectively.
We note that this observation holds for both nuclear configurations, indicating that the converged $\DD_\text{min}$ is more sensitive to the added noise than to the correlated characteristic of the underlying noiseless reduced state.

\begin{table}[H]
\centering
\caption{
Converged $\DD_{\min}$ from the ensemble ADAPT-VQA for 
the equilibrium and stretched $\mathrm{H_2}$ and linear $\mathrm{H_3}$ molecules. 
The targets are constructed by adding random noise of strength $\varepsilon$ to  the RDMs corresponding to canonical ensemble thermal states (see text for details). \label{tab:H2H3}}
\begin{tabular}{|c|>
{\centering\arraybackslash}p{2.5cm}|>{\centering\arraybackslash}p{2.5cm}|>{\centering\arraybackslash}p{2.5cm}|>{\centering\arraybackslash}p{2.5cm}|}
\hline
& \multicolumn{4}{c|}{$\DD_\text{min}$} \\
\cline{2-5}
 & \multicolumn{2}{c|}{$\mathrm{H}_2$} & \multicolumn{2}{c|}{$\mathrm{H}_3$}   \\
 \cline{2-5}
 & 1-RDM & 2-RDM & 1-RDM & 2-RDM  \\
\cline{2-5}
$\varepsilon$ & \multicolumn{4}{c|}{$\mathrm{R_{HH}}$=0.75~{\AA}}   \\
\cline{1-5}
$0.0$	    &  $0.0$   & $4.95 \times 10^{-9}$   & $4.45 \times 10^{-9}$      &    $1.98 \times 10^{-5}$  \\
$10^{-2}$	&  $3.11 \times 10^{-4}$	& $8.12 \times 10^{-3}$   & $1.08 \times 10^{-3}$	   &  $4.24 \times 10^{-2}$     \\
$10^{-1}$	&  $3.55 \times 10^{-2}$	& $7.71 \times 10^{-1}$   & $9.27 \times 10^{-2}$	   &  $4.08$ 	\\      
\cline{2-5}
& \multicolumn{4}{c|}{$\mathrm{R_{HH}}$=1.5~{\AA}}   \\
\cline{2-5}
 $0.0$	    &  $0.0$   &  $1.76 \times 10^{-7}$ &  $4.11 \times 10^{-10}$     &   $4.73 \times 10^{-5}$          \\
$10^{-2}$	&  $4.08 \times 10^{-4}$	&  $7.67 \times 10^{-3}$ &  $9.40 \times 10^{-4}$	   &  $4.44 \times 10^{-2}$                 \\
$10^{-1}$	&  $4.95 \times 10^{-2}$	&  $7.98 \times 10^{-1}$ &  $6.47 \times 10^{-2}$	   &  $4.20$ 	                       \\
\hline
\end{tabular}
\end{table}

\section{Summary}

We have proposed and assessed a practical framework for  determining the ensemble $N$-representability of an (alleged) $p$-body RDM. 
To this end,  
we employ a purification scheme that  embeds an ensemble state into a pure state defined on an extended Hilbert space, in a way that the reduced density matrices of the purified state reproduce those of the original ensemble.
We then  utilize a recently reported unitary-evolution algorithm for pure RDMs
to 
iteratively apply variational unitaries to an initial purified state. 
The  algorithm  effectively minimizes the distance between the corresponding $p$-body reduced density matrix and a specified target  $p$-body  matrix, alleged to be a proper ensemble $p$-RDM. 
This methodology can be used for both, correcting the lack of ensemble $N$-representability of target matrices and reconstructing a quantum-state from the corrected targets,
offering a route for $p$-RDM refinement. Moreover, the proposed algorithm is compatible with quantum computers.

We have validated the algorithm with numerical simulations on systems of two, three, and four electrons in both, model problems and molecular examples. 
For model systems of four electrons in three and four orbitals, we 
have assessed the effectiveness of the algorithm to decide 
on the pure or ensemble nature of 1-RDMs, and   
compared against Klyachko's  pure $N$-representability conditions.
For 2-RDMs, where the sufficient pure $N$-representability conditions are unknown, our  algorithm offers a practical route for dealing with this problem. 
We have also considered the capability of the ADAPT-VQA to deal with cases where the target matrices may not be $N$-representable. 
To this end, we have generated target matrices by   
artificially breaking the $N$-representability of 1- and 2-RDMs in a model system,  and in molecular \ch{H2} and \ch{H3} at finite temperature. 
Overall, the numerical results show that the  proposed algorithm is effectively able to detect the type of $N$-representability (pure or ensemble) of a target matrix, or the lack of it, as well as 
provide a closest RDM to a given non-$N$-representable target matrix and the corresponding (pure or purified) state.

\begin{suppinfo}
Klyachko's pure $N$-representability inequalities for the $(4e,3o)$ and $(4e,4o)$ model systems, and number of iterations required to achieve  convergence for the various systems in this work. 
\end{suppinfo}

\begin{acknowledgement}
OBO, GEM, PC, and DRA acknowledge the financial support from the Consejo Nacional de Investigaciones Cient\'{\i}ficas y T\'ecnicas (grant No. PIP KE3 11220200100467CO and PIP KE3 11220210100821CO). OBO, GEM, PC, and DRA acknowledge support from the Universidad de Buenos Aires (grant No. 20020190100214BA and 20020220100069BA)) and the Agencia Nacional de Promoci\'on Cient\'{\i}fica y Tecnol\'ogica (grant No. PICT-201-0381). JEP acknowledges support from NSF award number DMR-2318872. GES is a Welch Foundation Chair (C-0036), and his work was supported by the U.S. Department of Energy under Award No. DE-SC0019374. 
\end{acknowledgement}

\newpage

\bibliography{references}

\end{document}